\def\etal{{\it et al.\thinspace}}
\def\mearth{{\rm\,M_\oplus}}

\documentstyle[emulateapj,psfig,apjfonts]{article}

\received{}
\accepted{}
\journalid{}{}
\articleid{}{} 

\lefthead{}
\righthead{}

\begin{document}

\slugcomment{Short title: Terrestrial Accretion vs. Density Profile}

\title{Terrestrial Planet Formation in Disks with Varying Surface Density Profiles}

\author{Sean N. Raymond\altaffilmark{1}, Thomas Quinn\altaffilmark{1}, \&
Jonathan I. Lunine\altaffilmark{2}}

\altaffiltext{1}{Department of Astronomy, University of Washington, Box 351580,
Seattle, WA 98195 (raymond@astro.washington.edu; trq@astro.washington.edu)}
\altaffiltext{2}{Lunar and Planetary Laboratory, The University of Arizona,
Tucson, AZ 85287. (jlunine@lpl.arizona.edu)}

\begin{abstract}

The ``minimum-mass solar nebula'' (MMSN) model estimates the surface
density distribution of the protoplanetary disk by assuming the
planets to have formed {\it in situ}.  However, significant radial
migration of the giant planets likely occurred in the Solar system,
implying a distortion in the values derived by the MMSN method.  The
true density profiles of protoplanetary disks is therefore uncertain.
Here we present results of simulations of late-stage terrestrial
accretion, each starting from a disk of planetary embryos.  We assume
a power-law surface density profile that varies with heliocentric
distance $r$ as $r^{-\alpha}$, and vary $\alpha$ between 1/2 and 5/2
($\alpha = 3/2$ for the MMSN model).  We find that for steeper
profiles (higher values of $\alpha$), the terrestrial planets (i) are
more numerous, (ii) form more quickly, (iii) form closer to the star,
(iv) are more massive, (v) have higher iron contents, and (vi) have
lower water contents.  However, the possibility of forming potentially
habitable planets does not appear to vary strongly with $\alpha$.

 \end{abstract}



\section{Introduction}

In calculating the range of possible terrestrial planet systems that
could form around Sun-like stars, it is important to consider
protoplanetary disks with surface density profiles different that the
standard, ``minimum mass solar nebula'' model (MMSN -- e.g.,
Weidenschilling 1977). Different disk profiles form protoplanets with
different masses and spacings (Kokubo \& Ida 2002). The accretion of
terrestrial planets from these protoplanets is strongly affected by
the distribution of mass in the system, and the dynamical timescales
associated with regions of enhanced density. The terrestrial planets
that form in disks with varying density profiles may vary in
significant ways, e.g. in their compositions and formation
timescales. In this paper, we use a standard prescription for disk
mass densities, constrained by observations of such disks, to explore
how disk properties affect the growth of terrestrial planet systems.

The MMSN model calculates the density profile of the protoplanetary
disk in the following way.  The mass in each planet is augmented to
Solar composition by adding the appropriate amount of light elements.
Next, this mass is smeared out into contiguous annuli.  Finally, a
simple, power-law profile is fit to the corresponding data points.
The resulting surface density profile scales with heliocentric
distance $r$ (in AU) as $\Sigma (r) = \Sigma_1 \, r^{-3/2}$.  Values
for $\Sigma_1$ range from 1700 $g\,cm^{-2}$ (Hayashi 1981) to 4200
$g\,cm^{-2}$ (Weidenschilling 1977).  In this case, $\Sigma_1$
includes both the gas and dust component of the disk.  The solid
component of the MMSN follows the same $r^{-3/2}$ profile with
$\Sigma_1 \, \approx \, 6 \, g\,cm^{-2}$.

There is no reason to expect that all disks should follow the
standard, $r^{-3/2}$ MMSN profile.  There is evidence that at least
the outer planets in our Solar System may have undergone significant
radial migration (e.g. Levison \& Morbidelli 2003), which would
distort the inferred density profile.  In addition, certain disk
collapse models predict shallower, $r^{-1}$ profiles (Shakura \&
Sunyaev 1973).  In contrast, Kuchner (2004) showed that the mean,
minimum-mass extra solar nebula has a steeper, roughly $r^{-2}$
profile.  Of course, this steeper profile also assumes that the known
extra-solar giant planets formed and remained {\it in situ}.  In light
of studies of planet migration (e.g. Lin \etal 1996), this is not
thought to be the case.  There is clearly reason to consider
protoplanetary disks with surface density profiles different that the
MMSN model.

Wetherill (1996) formed planets in both $r^{-1}$ and $r^{-3/2}$
profiles, and saw no differences between the two cases.  Chambers \&
Cassen (2002) simulated the late-stage accretion of terrestrial
planets in two different surface density profiles.  One of these was a
simple, $r^{-1}$ profile, while the other had a low density in the
inner disk, with a peak at roughly 2 AU, based on a temperature
dependent condensation model of Cassen (2001).  They found that the
power law profile generally provided a better fit to the Solar system
than the peaked profile, in terms of the number and masses of
terrestrial planets formed.  There remains a great deal of uncertainty
in disk models in terms of redistribution of solids throughout the
accretion process, including gas drag, type I migration, and dynamical
scattering.

Raymond, Quinn \& Lunine (2004; hereafter RQL04) slightly varied the
value of $\Sigma_1$, but only between 8-10 $g\,cm^{-2}$.  They did,
however, vary the surface density in the outer protoplanetary disk by
changing the planetesimal mass used in the outer terrestrial zone.  We
found that a higher surface density results in the formation of a
smaller number of less massive planets.  These planets also tend to
have higher orbital eccentricities.  Raymond, Quinn \& Lunine (2005;
hereafter RQL05) ran two simulations with $\Sigma_1 = 2 g\, cm^{-2}$.
The systems that formed resembled extended asteroid belts, containing
several Mars-sized bodies but no large terrestrial planets.

We adopt a general power-law distribution $\Sigma (r) \, = \, \Sigma_1
\, r^{-\alpha}$ (note that $\alpha$ bears no relation to the
viscosity parameter by the same name).  We consider three different
power-law surface density distributions, with $\alpha$ = 0.5, 1.5 and
2.5.  These cases encompass the likely range of actual values,
according to theory and observation.  We determine the effects of
$\alpha$ on the accretion process as well as the physical properties
of the planets which form, including their mass, composition, and
potential habitability.

\section{Initial Conditions}

We consider three surface density distributions.  Each contains 10
Earth masses ($\mearth$) of material between 0.5 and 5 AU.  Because of
their different shapes, each has a different value for $\Sigma_1$.
For $\alpha$ = 0.5, $\Sigma_1 = \,5.7 g\,cm^{-2}$, for $\alpha$ = 1.5,
$\Sigma_1 = 13.5 \, g\,cm^{-2}$, and for $\alpha$ = 2.5, $\Sigma_1 =
21.3 \, g\,cm^{-2}$.  The mass contained in the inner disk is much
higher for higher values of $\alpha$.  The disk with $\alpha=2.5$ has
6.5 $\mearth$ of material inside 1.5 AU, compared with 3.6 $\mearth$
for $\alpha=1.5$ and only 1.4 $\mearth$ for $\alpha=0.5$.

We assume in our initial conditions that oligarchic growth has taken
place, and the disk mass is dominated by planetary embryos, shown to
be the case for different values of $\alpha$ by Kokubo \& Ida (2002).
Each planetary embryo has swept up the mass in its feeding zone, which
has a width of $\Delta$ mutual Hill radii.  With our generalized
power-law density profile, the mass of a given planetary embryo scales
with its semimajor axis $a$ and the separation between embryos
$\Delta$ as

\begin{equation}
M_{embryo} \propto a^{(2-\alpha)\, 3/2} \Delta^{3/2} \Sigma_{1}^{3/2}.
\end{equation}

\begin{figure}
\centerline{\plotone{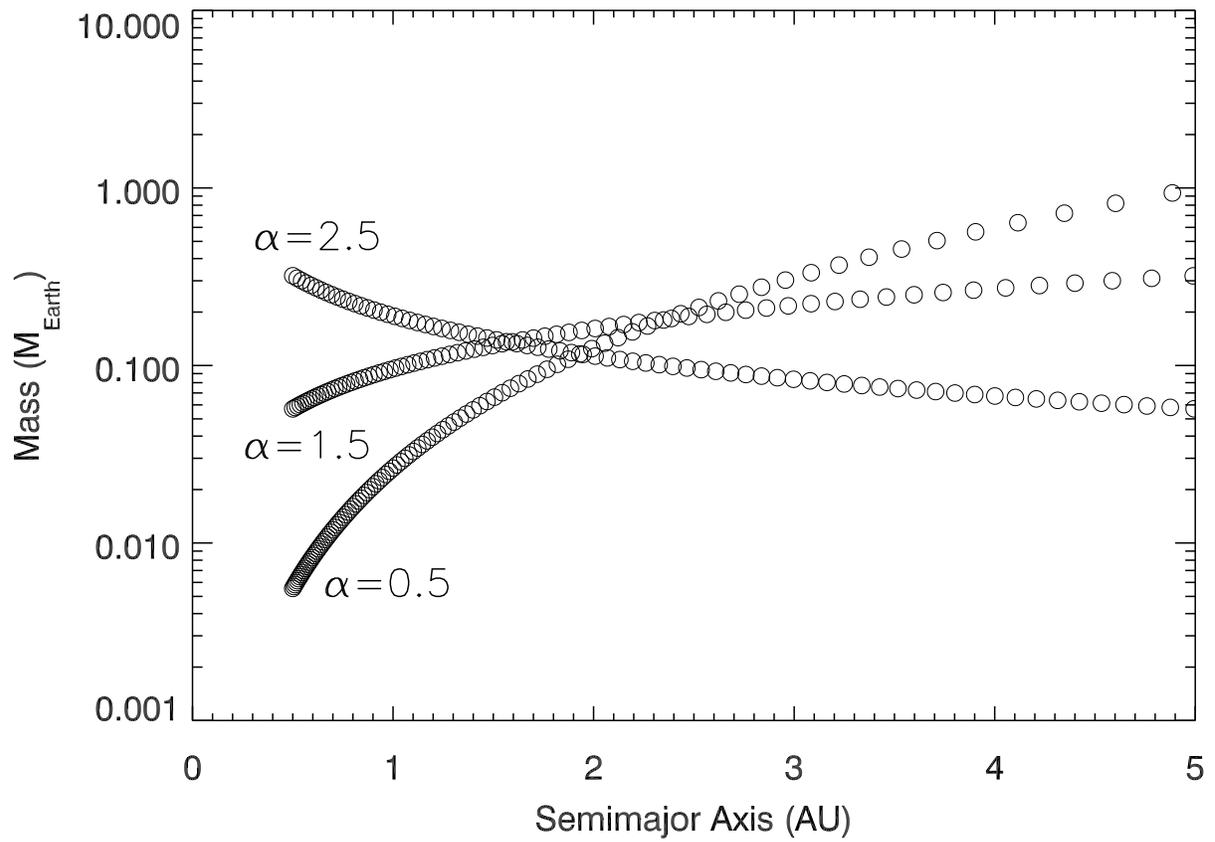}}
\caption{Distribution of planetary embryos for our three surface
  density profiles.  In this plot, embryos are spaced evenly by 7
  mutual Hill radii (i.e., $\Delta$ is fixed at 7).  In our actual
  simulations, $\Delta$ varies randomly between 1 and 2, increasing
  the number and decreasing the individual masses of embryos, as
  compared with this figure.}
\label{fig:a-init}
\end{figure}

For a density profile steeper than $\alpha$ = 2, the mass of planetary
embryos decreases with heliocentric distance, and the mass is very
concentrated close to the star.  Conversely, shallower density
profiles result in higher-mass planetary embryos at large distances,
with more mass in the outer disk.  Figure~\ref{fig:a-init} shows the
distribution of evenly spaced planetary embryos for each of our
surface density distributions.  For $\alpha = 0.5$, embryos are very
small in the inner disk and quite massive (up to $\mearth$) and
well-spaced in the outer disk.  Conversely, for $\alpha=2.5$, embryos
are smaller in the outer disk, though better-separated, as the Hill
radius has a stronger dependence on semimajor axis $a$ than on mass
$M$ ($R_H
\propto a\,M^{1/3}$).

In Fig.~\ref{fig:a-init}, $\Delta$ is fixed at 7 Hill radii, to
demonstrate the differences in embryo mass and spacing.  However, in
our simulations we space embryos randomly by 1-2 Hill radii from 0.5
to 5 AU.  The number of embryos varies for each value of $\alpha$: we
include roughly 380 embryos for $\alpha = 0.5$, 290 embryos for
$\alpha = 1.5$, and 280 embryos for $\alpha = 2.5$.  For each value of
$\alpha$, we perform three simulations with different random number
seeds.

The starting water contents of embryos are as described in RQL04:
embryos are dry inside 2 AU, contain 0.1\% water between 2-2.5 AU, and
5\% water past 2.5 AU.  Their initial iron contents are interpolated
between the values for the Solar System planets and chondritic
meteorites associated with known classes of primitive asteroids
(values from Lodders \& Fegley, Jr. 1999).  Embryos are given small
initial eccentricities ($e \leq 0.02$) and inclinations ($i \leq
1^{\circ}$).

In each case we include a Jupiter-mass giant planet at 5.5 AU on a
circular orbit.  One could argue that if the mass of a giant planet
scales with the isolation mass for a given surface density, then the
masses of the giant planets should vary in position and mass according
to our values of $\alpha$ (Lissauer 1995).  However, we do not know
which profile to be the most common in nature, or whether there even
exists a universal density profile for protoplanetary disks.  We are
constrained by observations of actual extra-solar planets which do
form, whatever their initial disk profile.  Due to the large amount of
uncertainty in this issue, it seems reasonable to assign the same
outer giant planet to each disk.

Each simulation was integrated for 200 Myr using the hybrid algorithm
of the Mercury integrator (Chambers 1999), with a 6 day timestep.
Collisions are treated as inelastic mergers which conserve linear
momentum, mass, water and iron content.  Each simulation took 1-3
months to run on a desktop PC, and conserved energy to at least one
part in $10^3$.

\section{Results}

\subsection{Details of two Simulations}

\begin{figure}
\centerline{\plotone{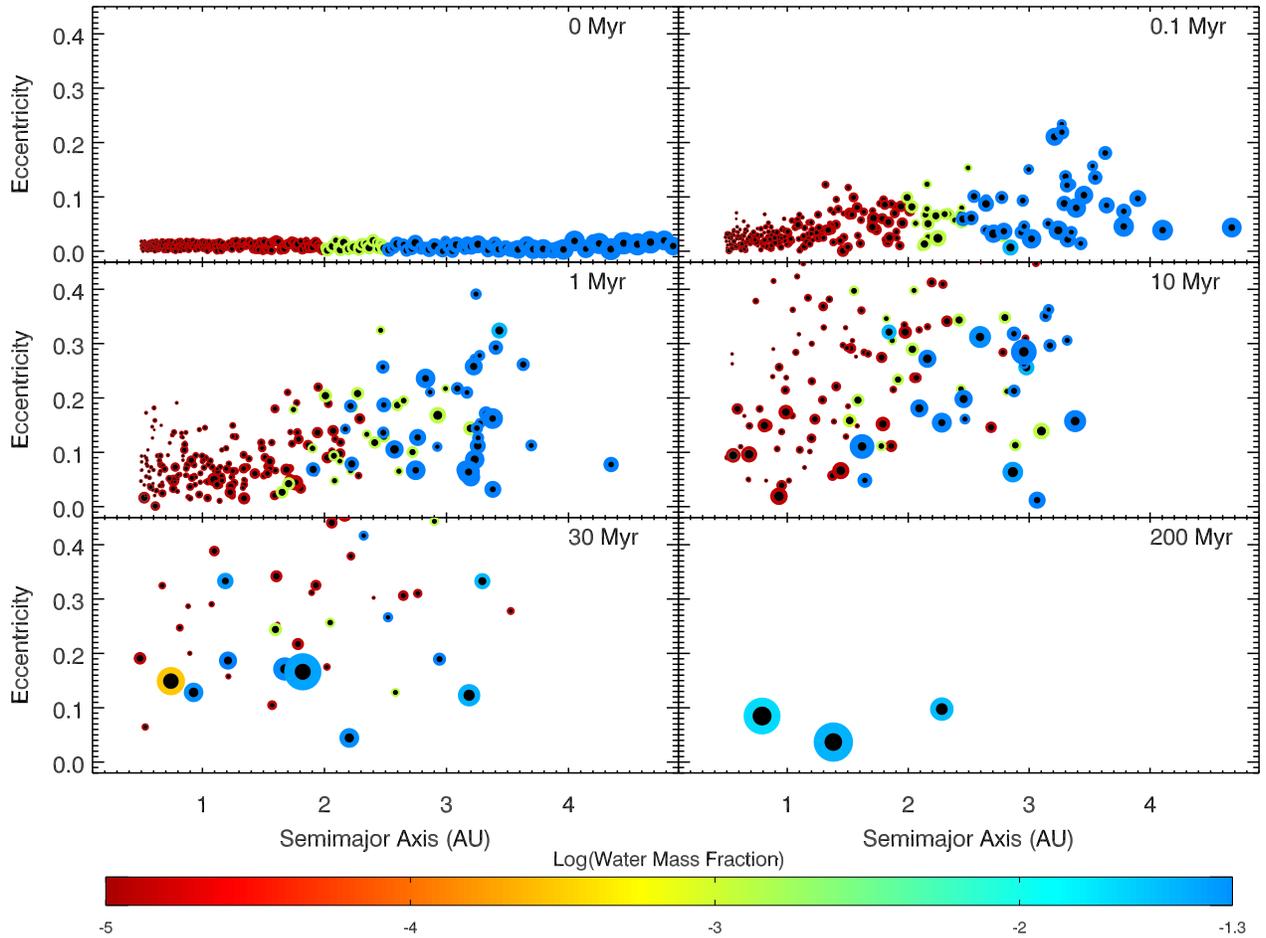}}
\caption{Six snapshots in time of the evolution of a simulation with
  $\alpha$ = 0.5, which formed two planets inside 2 AU.  The size of
  each body represents its relative physical size, but is not to
  scale.  The color represents its water content, and the dark circle
  in the center represents the size of its iron core.  The inner one
  has semimajor axis $a$ = 0.79 AU, eccentricity $e$ = 0.08, mass $M$
  = 1.4 $\mearth$, and a water mass fraction $W.M.F.$ of 1.6$\times
  10^{-2}$.  The second planet has $a$ = 1.38 AU, $e$ = 0.04, $M$ =
  1.7 $\mearth$, and $W.M.F.$ = 2.7 $\times 10^{-2}$.}
\label{fig:a-05a}
\end{figure}

\begin{figure}
\centerline{\plotone{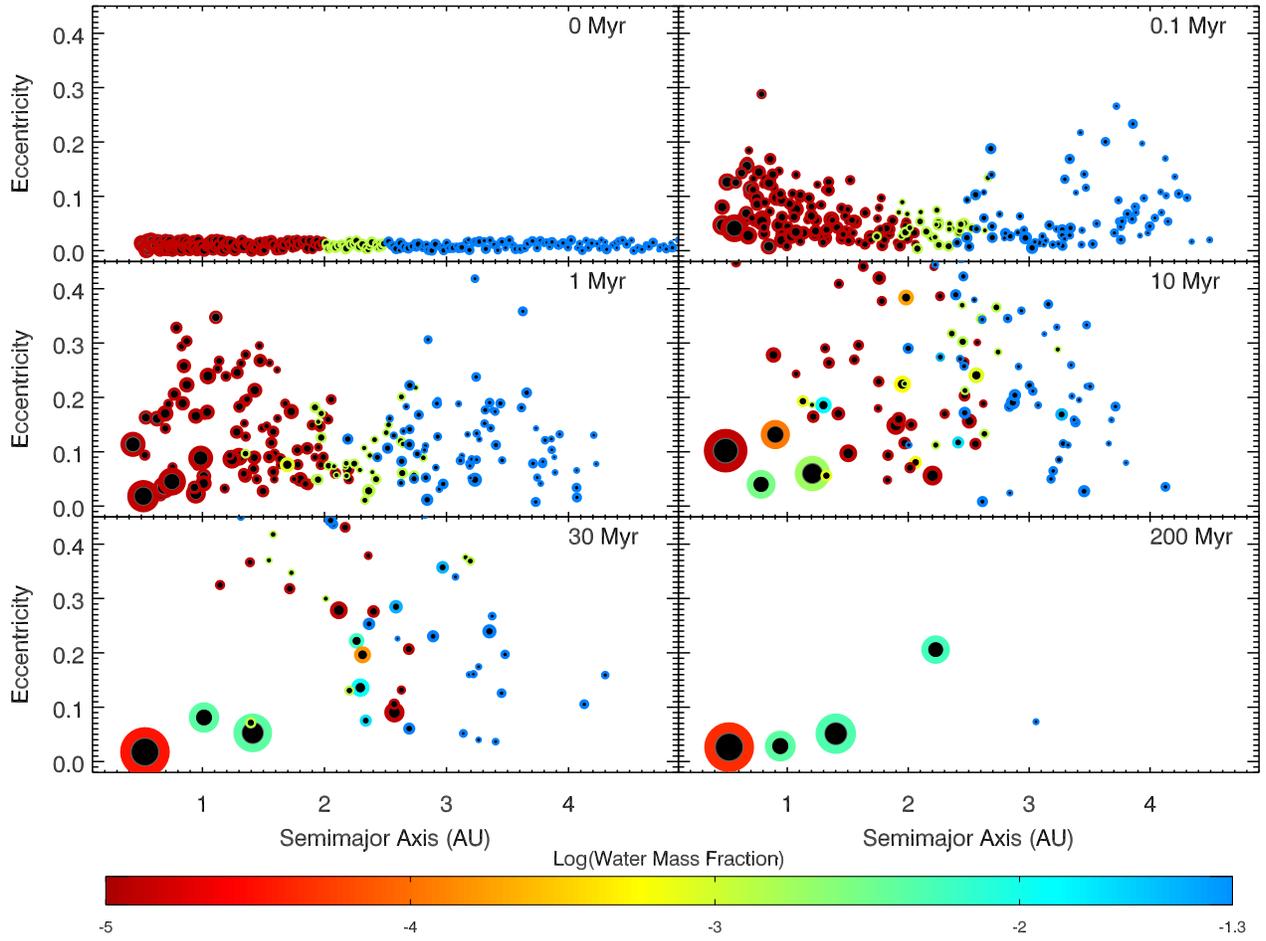}}
\caption{Six snapshots in time of the evolution of a simulation with
  $\alpha$ = 2.5, which formed three planets inside 2 AU, formatted as
  in Fig.~\ref{fig:a-05a}.  The inner planet has $a$ = 0.52 AU, $e$ =
  0.03, $M$ = 3.4 $\mearth$, and $W.M.F.$ = 2.3 $\times 10^{-5}$.  The
  second has $a$ = 0.94 AU, $e$ = 0.03, $M$ = 0.8 $\mearth$, and
  $W.M.F.$ = 3.1 $\times 10^{-3}$.  The third planet has $a$ = 1.40
  AU, $e$ = 0.05, $M$ = 1.8 $\mearth$, and $W.M.F.$ = 3.7 $\times
  10^{-3}$.}
\label{fig:a-25b}
\end{figure}

Figures~\ref{fig:a-05a} and~\ref{fig:a-25b} show the time evolution of
two simulations with $\alpha$ = 0.5 and 2.5, respectively.  The
overall progression of both simulations is similar, but the details
are quite different.  In both cases, the eccentricities of planetary
embryos increase until their orbits cross and collisions may occur.
This eccentricity pumping occurs due to secular and resonant
perturbations from both other embryos and the giant planet at 5.5 AU
(which, for simplicity, we now refer to as Jupiter).  Many embryos are
ejected after close encounters with Jupiter.  Larger bodies grow by
accreting smaller ones, on timescales which depend on both the
dynamical timescale (orbital distance) and the local density of
material.  Many bodies which form inside 2 AU contain one or more
water-rich embryos which began the simulation past 2-2.5 AU.  In time,
most small bodies are removed from the simulation, either via
collisions with the forming planets, the Sun, or Jupiter, or via
ejection from the system.  What remains are a few (typically 2-4)
terrestrial planets with significant masses, generally inside 2 AU.
The details of the planets which have formed in Figs.~\ref{fig:a-05a}
and~\ref{fig:a-25b} are summarized in the captions.

The main difference between the two simulations has to do with their
different mass distributions, and can be easily seen in a comparison
of the eccentricities of the inner disk in the 0.1 Myr panels of
Figs.~\ref{fig:a-05a} and~\ref{fig:a-25b}.  In~\ref{fig:a-05a}, where
$\alpha$ = 0.5, the majority of the terrestrial mass is in the outer
disk, past 2.5 AU.  Eccentricity pumping occurs mainly in the outer
disk, via both mutual encounters and interactions with Jupiter, and
the terrestrial planets therefore form from the outside in.  The
planets which form contain a large amount of water-rich material,
simply because so much initial material was found in the outer disk.
The final planets contain only a small fraction of the initial mass,
because embryos in the outer disk are very strongly excited by
Jupiter, and largely ejected from the system.

In Fig.~\ref{fig:a-25b}, where $\alpha$ = 2.5, the majority of the
mass lies inside 2.5 AU.  The eccentricities of embryos in the inner
disk are self-excited through mutual gravitational interactions.
Accretion proceeds quickly in the inner regions due to both the high
density of material and short dynamical timescales.  Terrestrial
planets therefore form from the inside out.  Most of the mass in the
system is well-separated from Jupiter, so the final planets contain
the majority of the initial mass.  However, since there was little
initial water-rich mass past 2.5 AU, they are much drier than in
Fig.~\ref{fig:a-05a}.

\subsection{Trends with $\alpha$}

\begin{figure}
\centerline{\plotone{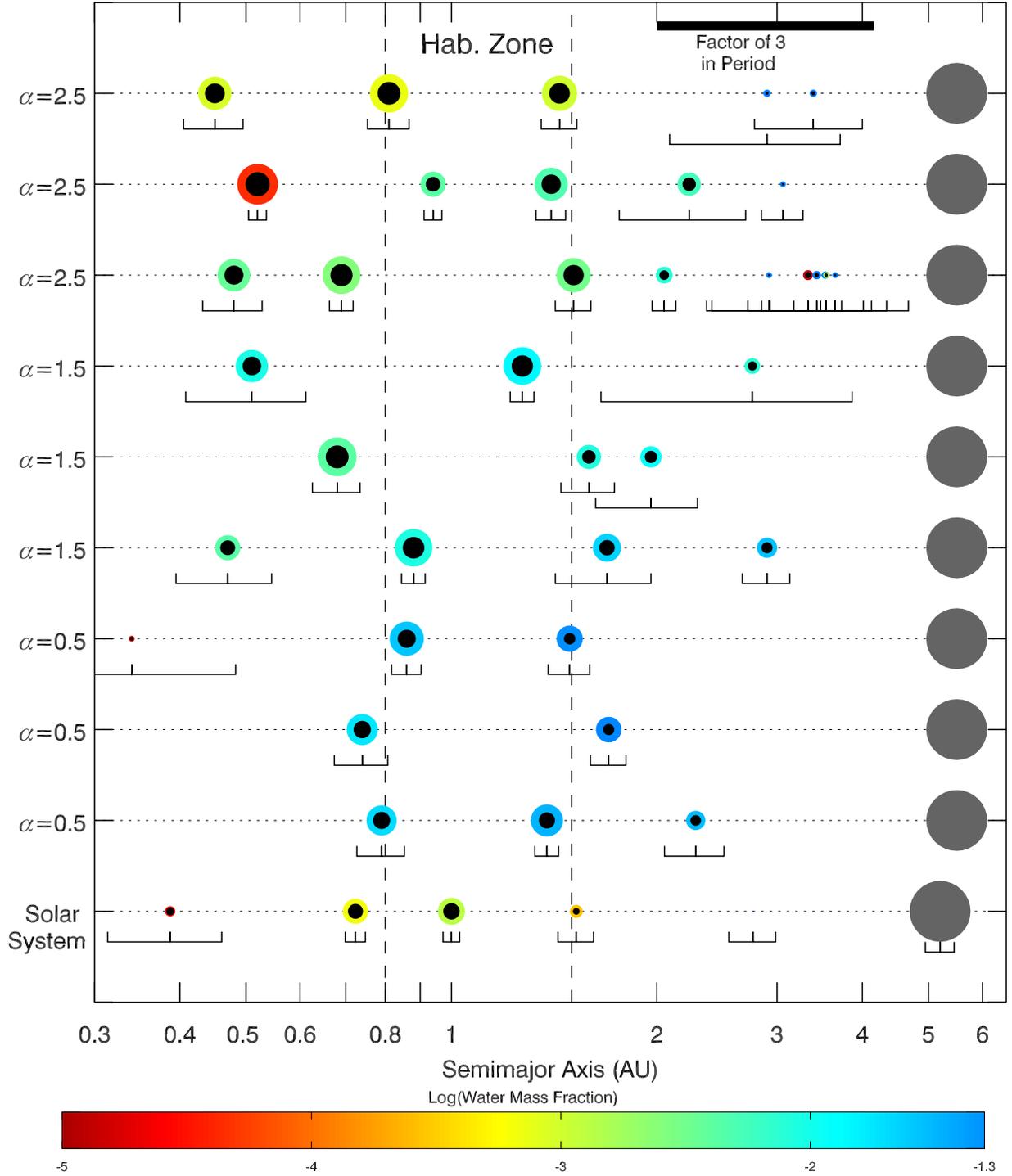}}
\caption{Final configuration of all 9 simulations, with the Solar
  System shown for comparison.  The size of each body corresponds to
  its relative physical size, but is not to scale on the x-axis.  The
  gray circles represent the location of the giant planets in each
  simulation, and are not on the same scale as the terrestrial bodies.
  The color of each planet represents its water content, and the dark
  circle in its center represents the size of its iron core.  The
  eccentricity of each body is shown beneath it, by its radial
  excursion over an orbit.}
\label{fig:a-final}
\end{figure}

Figure~\ref{fig:a-final} shows the final configuration of all nine
simulations we performed, with the Solar System shown for comparison.
Our simulations with $\alpha$ = 1.5 do not reproduce the Solar System
planets as well as previous simulations with the same profile, such as
RQL04.  This is because the total mass in the disk is higher in these
simulations.  The 10 $\mearth$ in the terrestrial region is roughly
twice as much as in the MMSN model, and $\sim$50\% higher than in
RQL04 (higher still for the case (ii) simulations).  As shown in
RQL04, a higher surface density of material results in the formation
of a larger number of more massive planets with larger eccentricities.
We see the results of this effect in comparing the $\alpha$ = 1.5
simulations with the Solar System in Fig.~\ref{fig:a-final}.  None of
our simulations form $\sim$Mars-sized planets inside 2 AU, also
because of the large total mass in protoplanets.  Although this
affects certain comparisons between our results and the Solar System
planets, it does not affect comparisons between simulations with
similar initial conditions.

There exist several strong correlations between the final planetary
systems and the slope of the density profile, $\alpha$, which are
shown in Table 1. Perhaps the most striking is simply the location
of the planets formed.  In cases with $\alpha$ = 1.5 and 2.5, the
innermost planet typically resides around 0.5 AU, while the innermost
planet in the $\alpha$ = 0.5 simulations is usually around 0.75 AU.
It is important to note that in this analysis we define a planet to be
a body larger than 0.2 $\mearth$ whose orbit lies inside 2 AU.

\begin{deluxetable}{l|ccc}
\tablecaption{Accretion Trends with $\alpha$}
\tablecolumns{4}
\tablehead{
\colhead{Trend} &  
\colhead{$\alpha=0.5$} & 
\colhead{$\alpha=1.5$} & 
\colhead{$\alpha=2.5$} }
\startdata
Mean innermost planet\tablenotemark{1}  & 0.74 AU & 0.47 AU & 0.45 AU \\
Mean planet mass & 1.4 $\mearth$ & 1.7 $\mearth$& 2.2 $\mearth$\\
Mean formation time\tablenotemark{2} & 55 Myr& 42 Myr& 18 Myr\\
Mass (planets) / Mass (disk)\tablenotemark{3} & 0.28 & 0.44 & 0.65 \\
Total mass in planets & 2.8 $\mearth$& 4.4 $\mearth$& 6.5 $\mearth$\\
Mean number of planets & 2.0 & 2.67 & 3.0 \\
Mean number of planets $<$ 1 AU & 1.0 & 1.3 & 2.0 \\
Avg. mass of planets $<$ 1 AU & 1.65 $\mearth$& 2.06 $\mearth$& 2.24 $\mearth$\\
Mean largest planet & 1.75 $\mearth$& 2.8 $\mearth$& 3.0 $\mearth$\\
Mean orbital eccentricity & 0.07 & 0.12 & 0.06\\
Mean water mass fraction & $2.6 \times 10^{-2}$ & $8.1 \times 10^{-3}$ &
$1.8 \times 10^{-3}$ \\
Mean iron mass fraction & 0.19 & 0.27 & 0.29 \\
Avg. num. of habitable planets\tablenotemark{4} & 1.0 & 2/3 & 2/3\\
\enddata

\tablenotetext{1}{We define a planets to be more massive than 0.2
  Earth masses and to have semimajor axis $a\,<$ 2 AU.}
\tablenotetext{2}{Mean time to reach 90\% of a planet's final
  mass. See Fig.~\ref{fig:a-tform}.}
\tablenotetext{3}{Fraction of initial disk mass contained in final planets}
\tablenotetext{4}{Habitable planets are defined to have semimajor axes
  between 0.8 and 1.5 AU, and water mass fractions larger than $10^{-3}$.}
\end{deluxetable}

\begin{figure}
\centerline{\plotone{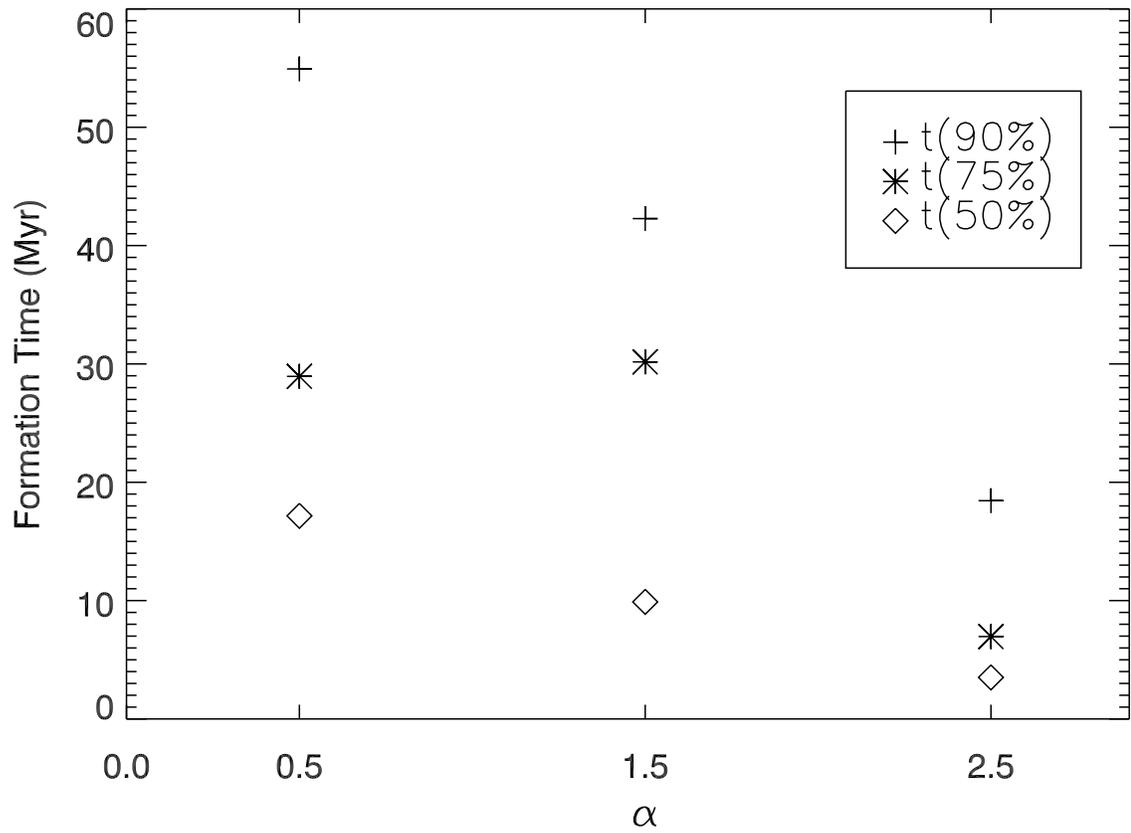}}
\caption{Timescales for planets to reach a given fraction (50\%, 75\%,
  or 90\%) of their final mass, as a function of surface density
  profile, $\alpha$.  These values are averages for all planets with
  $M > 0.2 \mearth$ and $a < 2$ AU.}
\label{fig:a-tform}
\end{figure}

The trends from Table 1 can be summarized as follows.  As compared
with a protoplanetary disk with a shallow surface density profile, a
disk with a steeper surface density profile forms a larger number of
more massive terrestrial planets in a shorter time.
Figure~\ref{fig:a-tform} shows the mean time for planets to reach
50\%, 75\%, and 90\% of their final mass for all three values of
$\alpha$.  The reason that $\alpha$ = 0.5 planets form more slowly is
that the timescale for accretion scales with the dynamical time and
the local density.  The density in the outer disk is high enough that
planets form more from the outside in.  The long dynamical timescales
translate to slower planet growth, and planets at larger heliocentric
distances.  In the case of $\alpha$ = 2.5 planets, the high inner
density and fast dynamical timescales conspire to form planets on
remarkably short timescales.  There is a constraint from measured Hf-W
ratios that both the Moon and the Earth's core were formed by $t\,
\approx$ 30 Myr (Kleine \etal, 2002; Yin \etal, 2002), suggesting that
the Earth was at least 50\% of its final mass by that time.
Fig.~\ref{fig:a-tform} shows that planets in each of our density
profiles satisfy this constraint.

A larger number of planets is formed in the simulations with $\alpha$
= 2.5 because the planets in those simulations occupy a larger range
in semimajor axes, i.e. they extend to smaller orbital radii than for
$\alpha$ = 0.5.  The mean orbital radii of the innermost planet for
$\alpha$ = 2.5 and $\alpha$ = 0.5 simulations are 0.45 AU and 0.74 AU,
respectively.  This corresponds to a difference in orbital period of
over a factor of two, providing enough dynamical room for another
planet.  Indeed, the $\alpha$ = 2.5 simulations formed an average of
three planets per simulation, as compared with two planets per
simulation for $\alpha$ = 0.5.

The reason that more mass ends up in planets for steeper density
profiles is the dynamical presence of Jupiter.  During accretion, a
large fraction of the material close to Jupiter is ejected from the
planetary system.  The ``ejection range'' of Jupiter, the zone from
which a large fraction of material is ejected, extends to a factor of
4-6 in orbital period (a factor of 2.5-3.3 in semimajor axis) from
Jupiter.  If we na\"ively assume as in RQL05 that all material within
a factor $F$ in orbital period of Jupiter is ejected from the system,
then the total mass ejected is

\begin{equation}
M_{ejec} = \frac{2 \pi \Sigma_1}{2-\alpha} \left[(r_J \,
F^{2/3})^{2-\alpha} - (r_J F^{-2/3})^{2-\alpha}\right],
\end{equation}

\noindent where $r_J$ is the orbital radius of Jupiter.  The value of
$F$ depends on the mass and eccentricity of the giant planet, and may
also scale with $r_J$ and $\alpha$.  If we only consider material
ejected from interior to the giant planet, then

\begin{equation}
M_{ejec} = \frac{2 \pi \Sigma_1}{2-\alpha}\, r_J^{2-\alpha} \left(1 -
F^{2/3 (\alpha-2)}\right).
\end{equation}

We know the values of $\Sigma_1$ and the mean amount of mass ejected
per simulation for each value of $\alpha$, so we can calculate a value
for $F$ (we also take into account the fact that embryos exist only
out to 5 AU, while $r_J$ = 5.5 AU).  We find that $F$ ranges between
roughly 4 ($\alpha= 0.5$) and 6 ($\alpha = 2.5$).  The mean location
of the outermost planet in each simulation, independent of $\alpha$,
was at about the 5:1 resonance with Jupiter (at 1.9 AU), consistent
with our values of $F$.  The range at which Jupiter ejects material
from an $\alpha$ = 2.5 disk appears to be farther than for an $\alpha$
= 0.5 disk (Jupiter's effective ``ejection range'' is the 6:1 vs 4:1
resonance for $\alpha$ = 2.5 and 0.5, respectively).  The reason for
this increased range for $\alpha$ = 2.5 is the lack of material in the
outer disk with which excited embryos may interact, possibly damp
their eccentricities, and remain in the disk.  A particle excited by
Jupiter in an $\alpha$ = 0.5 disk may have its eccentricity
subsequently decreased via embryo-embryo interactions, whereas the
dearth of material in the outer disk prevents this from happening for
$\alpha$ = 2.5.

We perform a MMSN-like test on the planets which formed.  For each
system, we select all planets of 0.2 $\mearth$ or more and spread
their mass into concentric, nested annuli, and calculate an effective
surface density.  We choose the division of annuli between two planets
to be the geometric mean of the semimajor axes of the two planets,
following Kuchner (2004).  For $\alpha = 2.5$, we find a best-fit
slope to the surface density, $\alpha_{fit}$, of almost exactly 2.5.
However, we obtain $\alpha_{fit}$ values between 2.1 and 2.3 for the
planets formed in $\alpha$ = 0.5 or 1.5 simulations.  The reason for
this is because, as discussed above, Jupiter's ejection of material
from the asteroid region has a much stronger effect on shallower
density profiles.  This decreases the surface density measurements for
outer radii, skewing the fit to a steeper profile.  If we only include
planets inside 2 AU, outside of Jupiter's ejection range, the
$\alpha_{fit}$ for the $\alpha$ = 1.5 planets changes to 1.53.
However, $\alpha_{fit}$ for the $\alpha$ = 0.5 planets remains above
2, because of Jupiter's pervasive effects in low-$\alpha$ simulations.
Choosing an even smaller outer radius would likely yield a value
closer to 0.5, but the number of planets is too small to perform the
test.  This analysis implies that, in the absence of external
perturbations (e.g. from the giant planets), terrestrial accretion
forms planets that conform to their initial density profile.  In this
sense, the planets ``remember'' the initial state of the disk.

This begs the question of why the Solar system planets follow an
$r^{-1.5}$ profile.  Two explanations are possible: 1) the surface
density profile of the protoplanetary disk followed an $r^{-1.5}$
profile, or 2) the initial profile was different, but perturbations
caused the final planets to follow an altered profile.  If
perturbations did affect the final planets, then the starting mass
profile was likely flatter, i.e., the starting value of $\alpha$ was
smaller than 1.5.  However, as demonstrated above, the ejection range
of Jupiter extends inward to orbital radii of $\sim$ 2 AU.  Additional
secular perturbations arise from the presence of Saturn (e.g. the
$\nu_6$ secular resonance at 2.1 AU).  But the inner terrestrial
region remains relatively sheltered from these perturbations.  In
addition, the giant planets' eccentricities may have been
significantly lower during the epoch of terrestrial accretion
(Tsiganis \etal 2005), reducing the effect of secular resonances.  We
conclude that the inner solar nebula probably did follow an $r^{-1.5}$
profile.


The probability of a habitable planet forming between 0.8-1.5 AU with
water mass fraction $> 10^{-3}$ does not depend strongly on the value
of $\alpha$, although planets that form in disks with different
$\alpha$ values will certainly have vastly different characteristics.
Planets that form with $\alpha$ = 2.5 have much higher iron contents
than those in $\alpha$ = 0.5 disks, simply because the amount of
iron-rich material close to the star is much higher.  Similarly,
planets that form in $\alpha$ = 0.5 disks have much higher water
contents than those in $\alpha$ = 1.5 or 2.5 disks, because the bulk
of the solid mass in $\alpha$ = 0.5 disks is found in the water-rich
outer regions.  The iron and water contents will have large
consequences for the nature of the surface and interior of these
planets, as well as their potential habitability.

\section{Discussion}

The true surface density profiles of the solid component of
protoplanetary disks are unclear.  Their formation involves a great
deal of physics, including the condensation sequence as a function of
temperature, and radial migration via gas drag and type I migration.
It is likely that the standard, minimum mass solar nebula model is
oversimplified, as there is strong evidence from the dynamical
structure of the Kuiper belt that the giant planets experienced
significant radial migration (Levison \& Morbidelli 2003).

We have demonstrated that the terrestrial planets which form in
protoplanetary disks with three different power-law surface density
profiles have very different characteristics.  The total number and
mass of terrestrial planets is strongly affected by the surface
density slope, $\alpha$.  In the presence of an outer giant planet, a
disk with a steeper profile (higher $\alpha$) yields a larger number
of more massive planets than a shallower profile.  Planets in
steeper-profiled disks form more quickly, have higher iron content and
lower water content than for shallower profiles, with implications for
the habitability of planets in each case.

It is unclear whether our placement of a Jupiter-mass giant planet at
5.5 AU is self-consistent for each chosen value of $\alpha$.  Kokubo
\& Ida (2002) and Lissauer (1995) argued that the character of the
giant planets may depend on the disk's density profile.  We chose
identical initial conditions to explore the effects of $\alpha$ in a
``typical'' planetary system, although the nature of such a system may
be inextricably linked to $\alpha$.  Some of our results that were
strongly influenced by the presence of a giant planet, such as the
strong depletion of the $\alpha=0.5$ disk, may only hold under our
given assumption of a giant planet at 5.5 AU.  However, many of our
results are independent of this assumption, such as the formation
timescales and compositions of the planets we form.

Using previous results, we can extrapolate the effects of other
parameters on our simulations, such as giant planet orbital radius and
eccentricity, and the location of the snow line.  It has been shown
that a giant planet on an eccentric orbit forms terrestrial planets
which are drier, less massive, and have larger eccentricities than for
a giant planet on a circular orbit (Chambers \& Cassen, 2002; RQL04).
In the presence of an eccentric giant planet, more of the material
from the asteroid region is ejected, and planets form at a larger
distance from the giant.  Therefore, an eccentric giant planet would
cause even more disparity between disks with shallow vs. steep density
profiles.  A disk with $\alpha = 0.5$ would have an even higher
fraction of its total mass ejected, causing the terrestrial planets to
be smaller.  An $\alpha = 2.5$ disk, however, would not be strongly
affected, since its mass is so centrally concentrated.  The primary
consequence of the giant planets' orbital radius is the amount of
solid material destroyed through ejection from the system.  This will
vary with heliocentric distance $r$ and $\alpha$ as shown in equation
2.  The location of the snow line determines how much water-rich
material exists in disks with different values of $\alpha$.

Kuchner (2004) found that applying the MMSN technique to the known,
extra-solar planets yielded a very steep density profile with $\alpha
\, \approx \, 2$.  This steep profile corresponds to an {\it in situ}
formation scenario for close-in giant planets (e.g. ``hot jupiters'').
If such a profile is common, then perhaps hot jupiters exist in these
systems.  We expect that terrestrial planets may form in the presence
of a hot jupiter, even if it has migrated through the terrestrial
region (RQL05).  Indeed, terrestrial planets
appear to be able to form in some of the known extra solar planetary
systems, such as 55 Cancri (Raymond \& Barnes 2005).

Real protoplanetary disks are not perfectly smooth power laws.  The
physics and chemistry determining their true shape is very complex,
and has not been fully modeled.  Some preliminary models show a peak
in surface density around 1-2 AU (Cassen 2001), which may explain the
large masses of Venus and Earth (Chambers \& Cassen 2002).  Such a
profile can be roughly approximated as having a small $\alpha$ value
in the inner disk, and a larger $\alpha$ value past $\sim$ 2 AU.  We
have shown that accretion proceeds from the inner disk outwards for
large $\alpha$ and from the outer disk inwards for smaller $\alpha$
values.  In this composite disk, accretion would likely start in the
highest density region around 1-2 AU, and proceed both inward and
outward.

\section{Acknowledgments}

We thank referee John Chambers for bringing up some important points.
This work was funded by grants from NASA Astrobiology Institute and
NASA Planetary Atmospheres.  These simulations were run under
Condor\footnote{Condor is publicly available at
http://www.cs.wisc.edu/condor}.

\end{document}